\documentclass[twocolumn,showpacs,preprintnumbers,amsmath,amssymb]{revtex4}

\usepackage{graphicx}
\usepackage{dcolumn}
\usepackage{bm}

\begin{document}


\title{On the mass of the $D_s(0^+,1^+)$ system}
\author{A. Deandrea}
\email{deandrea@ipnl.in2p3.fr}
\affiliation{Institut de Physique Nucl\'eaire, Universit\'e Lyon I, 4 rue
E.~Fermi,  F-69622 Villeurbanne Cedex, France}
\author{G. Nardulli}
\email{giuseppe.nardulli@ba.infn.it}
\affiliation{Dipartimento Interateneo 
di Fisica, Universit\`a di Bari and I.N.F.N.,
Sezione di Bari, I-70126 Bari, Italia}
\author{A.D. Polosa}
\email{antonio.polosa@cern.ch}
\affiliation{CERN - Theory Division, CH-1211 Geneva 23, Switzerland}

\date{July 1, 2003}

\begin{abstract}
In this note we discuss a determination for the mass of the
$D_s(0^+,1^+)$ system recently discovered by the BaBar, CLEO II
and Belle Collaborations. The value of the mass is derived by
making explicit the prediction obtained in a quark-meson model
prior to the discovery of these states.

\end{abstract}

\maketitle

Recently the BaBar collaboration \cite{Aubert:2003fg} discovered a
narrow meson near 2.32~GeV$/c^2$~decaying into $D_s\pi^0$. The
reported mass is
\begin{equation}M= 2316.8 \pm 0.4\, {\rm MeV/}c^2\ ,\label{baba}\end{equation}
while the width $\Gamma=8.6\pm 0.4$ MeV/$c^2$ is consistent with
the experimental resolution. This result has been confirmed by the
CLEO II Collaboration \cite{Besson:2003cp} and the Belle
Collaboration \cite{belle}. Besides the 2.32 state CLEO  also
finds another narrow state, near 2.46~GeV$/c^2$,~decaying into
$D^*_s\pi^0$. More precisely for the latter mass they find
\begin{equation}M'= 2463 \,{\rm MeV}/c^2\ .\label{cleo}\end{equation}
These data have triggered a discussion on the nature of the
observed states, see e.g.
\cite{Barnes:2003dj,Szczepaniak:2003vy,Nussinov:2003uj,Datta:2003iy,Terasaki:2003qa,Dai:2003yg,Bardeen:2003kt,vanBeveren:2003hj,vanBeveren:2003kd,Godfrey:2003kg,Cheng:2003kg,Cahn:2003cw,Colangelo:2003vg}
and
 the review
in~\cite{alex}. The natural interpretation should be that these
states form the  $J^P=(0^+,1^+)$ doublet,  predicted by the Heavy
Quark Effective Theory (HQET) \cite{Manohar:dt} and generally
denoted as $S$ in this formalism. Their masses happen to be below
the threshold mass for the decay into the Zweig allowed final
state $D^{(*)}K$ and this forces the isospin violating
$D_s^{(*)}\pi^0$ decay channel and the narrowness of the decaying
meson. Though straightforward, this identification is questioned
by some results of potential models
\cite{Godfrey:xj,Godfrey:wj,DiPierro:2001uu} that predict larger
masses, above the $D^{(*)}K$ threshold. Because of it, more exotic
explanations have been proposed in terms of baryonium, $D\pi$
atoms or $DK$ molecules
\cite{Barnes:2003dj,Szczepaniak:2003vy,Nussinov:2003uj,Datta:2003iy,Terasaki:2003qa}.

 Our point of view is very close to the one
expressed in \cite{Bardeen:2003kt} (see also
\cite{Bardeen:1993ae}). We do not see anything exotic in this
state, especially so because the mass $M_S^{(s)}$ of the
$S=(0^+,1^+)$ multiplet in the strange sector can be obtained
using a relativistic quark model incorporating the symmetries of
the HQET~\cite{noi}. The calculation is entirely analogous to the
one of the non-strange multiplet $(0^+,1^+)$~ whose mass was
predicted in \cite{noi}:\begin{equation}M_S=2165\pm 50\, {\rm
MeV}/c^2 \ .\label{ms}
\end{equation} A rough estimate of the strange
$S-$doublet mass  can be simply obtained by adding  a mass term
\begin{equation} M(D_s)-M(D)\simeq M(D^*_s)-M(D^*)\simeq 100\, {\rm MeV}/c^2\label{ms1}
\end{equation} to
(\ref{ms}). This would give a result $M^{(s)}_S\simeq 2265$
MeV/$c^2$, which would be compatible with (\ref{baba}) and
(\ref{cleo}) only if the theoretical uncertainties were larger.
 This estimate is however too crude, and in any case unnecessary
because the model allows a more
 precise calculation. In view of its interest we
present it below. It is worth noticing that not only the average
mass (\ref{ms}) of the $J^P=(0^+,1^+)$ non-strange states was
computed, but also the masses  of the strange
$J^P=(0^+,1^+)$ states were already evaluated by us. This
calculation is contained in \cite{noi0}  and its result was used
as a parameter in a form factor parametrization. In this note we
briefly recall the discussion outlined in~\cite{noi0} and make
explicit the prediction for $M_S^{(s)}$.

The model we consider is a quark-meson model (CQM), introduced in
\cite{noi} as an extension of the ideas and the methods in Refs.
\cite{Bardeen:1993ae}, \cite{manohar} and \cite{cqm}. A survey of
these topics can be found in \cite{nc}. The transition amplitudes
containing light/heavy mesons in the initial and final states as
well as the couplings of the heavy mesons to hadronic currents can
be calculated via quark loop diagrams where mesons enter as
external legs. The model is relativistic and incorporates, besides
the heavy quark symmetries, also the chiral symmetry of the light
quark sector.

The model can be extended to the strange quark sector solving
the gap equation discussed in \cite{ebert} with a non zero
current mass for the strange quark:
\begin{equation}
\Pi (m)=m - m_0 - 8\, m\, G\, I_1(m^2) = 0,
\end{equation}
where $G=5.25\;{\rm GeV}^{-2}$ and $m_0$ is the current mass of
the strange quark. The $I_1$ integral is calculated using the
proper time regularization:
\begin{equation}
I_1=\frac{iN_c}{16\pi^4} \int^{\mathrm{reg}} \frac{d^4k}{(k^2 - m^2)}
={\frac{N_c m^2}{16 \pi^2}} \Gamma\left(-1,\frac{m^2}{\Lambda^2},
\frac{m^2}{\mu^2} \right).
\end{equation}
The choice of the ultraviolet (UV) cutoff is dictated by the scale
of chiral symmetry breaking $\Lambda_\chi= 4 \pi f_\pi$ and we
adopted $\Lambda=1.25$ GeV. The infrared (IR) cutoff $\mu$ and the
constituent mass $m$ must be fixed taking into account that CQM
does not incorporate confinement. This means that one has to
enforce the kinematical condition to produce  free constituent
quarks $M \geq m_Q + m$, where $M$ is the mass of the heavy meson
and $m_Q$ is the constituent mass of the heavy quark. The heavy
meson momentum is $P^\mu=m_Q v^\mu + k^\mu$, $v^\mu$ being the
heavy quark 4-velocity and $k^\mu$ the so called residual momentum
due to the interactions of the heavy quark with the light degrees
of freedom at the scale of $\Lambda_{\rm QCD}$. Therefore the
above condition coincides with $v\cdot k \geq m$. This is so
because the 4-velocity of the meson is almost coincident with that
of the heavy quark, i.e. $P\simeq Mv$. Equivalently, in the rest
frame of the meson, ${\rm inf}(k)=m$, meaning that the smallest
residual momenta that can run in the CQM loop amplitudes are of
the same size of the light constituent mass. The IR cutoff $\mu$
is therefore $\mu \simeq m$.

A reasonable constituent quark mass for the strange quark is
$m=510$ MeV$/c^2$, considering the $\phi$ meson as a pure $s
\bar{s}$ state. Taking $\mu=0.51$ GeV$/c^2$ as an infrared cutoff,
a value of $m_0=131$ MeV$/c^2$  is required by the gap equation
(consistently with the spread of values for the current $s$ quark
mass quoted in \cite{pdg}). Varying the current strange mass in
the range $60-170$ MeV$/c^2$ gives a small excursion of
the constituent strange mass around the $500$ MeV$/c^2$ value.

The free parameter of CQM is $\Delta_H$ defined, in the infinite
heavy quark mass limit, by $\Delta_H=M_H-m_Q$. The subscript $H$
refers to the $H-$multiplet of the HQET \cite{Manohar:dt}
$H=(0^-,1^-)$. In a similar way a $\Delta_S$ is associated to the
$S$ multiplet, $S=(0^+,1^+)$. The latter is determined by fixing
$\Delta_H$ and solving the equation:
\begin{equation}
\Pi(\Delta_H)=\Pi(\Delta_S),
\label{eq:ric}
\end{equation}
where
\begin{equation}
\Pi(\Delta_{H,S})=I_1+(\Delta_{H,S}\pm m)I_3(\Delta_{H,S})
\end{equation}
with
\begin{eqnarray}
I_3(\Delta) &=& - \frac{iN_c}{16\pi^4} \int^{\mathrm {reg}}
\frac{d^4k}{(k^2-m^2)(v\cdot k + \Delta + i\epsilon)} \\
&=&\frac{N_c}{16{{\pi}^{{3/2}}}} \int_{1/{{\Lambda}^2}}^{1/{{\mu
}^2}} \frac{ds\,e^{- s( {m^2} - {{\Delta }^2} ) }}{s^{3/2}} \left(
1 + {\mathrm {erf}} (\Delta\sqrt{s}) \right)\, .\nonumber
\end{eqnarray}
Eq.~(\ref{eq:ric}) comes from requiring the HQET form of the kinetic 
term in the effective Lagrangian defining the model.

The related $\Delta^{(s)}_H,\Delta^{(s)}_S$ values in the strange
sector are shown in Table I. We consider in the table the range of
values $\Delta^{(s)}_H=0.5,0.6,0.7$ GeV$/c^2$, which is consistent
with the condition $M \geq m_Q+m$. Note that in the non strange
sector we considered in \cite{noi} the values
$\Delta_H=0.3,0.4,0.5$ GeV$/c^2$, smaller or higher values being
excluded by consistency argument or by experiment. Using
(\ref{ms1}) we see that the value $\Delta^{(s)}_H=0.7$ GeV$/c^2$
has to be excluded as well. \vskip1.cm\noindent
\begin{table}[ht]
\hfil \vbox{\offinterlineskip \halign{&#&
\strut\quad#\hfil\quad\cr \hline \hline &$\Delta^{(s)}_H$ &&
$\Delta^{(s)}_S$& \cr \hline &$0.5$&& $0.86$ &\cr &$0.6$&& $0.91$
&\cr &$0.7$&& $0.97$ &\cr \hline \hline }} \caption{
$\Delta^{(s)}$ (in GeV$/c^2$).} \label{t:senza1}
\end{table}

In order to give the explicit value of the mass of the $(0^+,1^+)$
states we observe that experimentally one has:
\begin{equation}
M^{(s)}_{H}=\frac{3 M_{D_s^*}+M_{D_s}}{4} = 2076 \pm 1~{\rm
MeV}/c^2.
\end{equation}
Considering only the first two entries in Table I, from
$\Delta^{(s)}_S-\Delta^{(s)}_H\equiv M^{(s)}_{S}-M^{(s)}_{H}=
335\pm 25$~MeV$/c^2$ one gets
\begin{equation}
M^{(s)}_{S}=2411\pm 25~{\rm MeV}/c^2 \label{11}
\end{equation} that differs by $\sim $140 MeV$/c^2$ from the rough
estimate presented above.

Eq. (\ref{11}) represents the main result of this note. It gives
the average mass of the $S=(0^+,1^+)$ doublet and is related to
the masses of the two states by
\begin{equation}
M^{(s)}_{S}=\frac{3 M_{D_s^*(1^+)}+M_{D_s(0^+)}}{4}\ .\label{ms2}
\end{equation}
From the measured value of the $0^+$ state, eq. (\ref{baba}), we
get 
\begin{equation}M_{D_s^*(1^+)}=\frac {
4\,M^{(s)}_{S}}3\,-\,\frac{M_{D_s(0^+)}}3\,=\,2442\pm 33\ {\rm
MeV}/c^2 ,
\end{equation} 
which agrees, within the theoretical
uncertainties, with the CLEO result (\ref{cleo}). The 
overall agreement is
better here than in other approaches, see e.g. the discussion
contained in~\cite{Cahn:2003cw}.

We note that in the present model the relation
$(M_{1^+}-M_{0^+})=(M_{1^-}-M_{0^-})\simeq 142$ MeV$/c^2$
\cite{Bardeen:2003kt} does not hold necessarily (indeed we find
$M_{1^+}-M_{0^+}=125$ MeV$/c^2$). In any case, assuming its
validity, from (\ref{ms}) we would get $M_{D_s^*(0^+)}=2304\pm 25$
MeV$/c^2$ and $M_{D_s^*(1^+)}=2446\pm 25$ MeV$/c^2$, which is also
compatible, within the errors, with the data (\ref{baba}) and
(\ref{cleo}). Let us finally observe that, using Table
I, we  expect for the $B_s(0^+,1^+)$ system a central mass of
$M=5740\pm 25$~MeV$/c^2$ and, for the two individual states,
$M_{B_s^*(0^+)}=5710\pm 25$~MeV$/c^2$ and $M_{B_s^*(1^+)}=5770\pm
25$~MeV$/c^2$ respectively. On the basis of these results, the
$\bar b s$ signal at 5850 MeV$/c^2$ \cite{pdg} should be better
interpreted as arising from the $(1^+,2^+)$ doublet predicted by
the HQET.

\acknowledgements{We thank A. Pompili for most useful
discussions. ADP is supported by a M.~Curie fellowship, contract 
HPMF-CT-2001-01178.}

\end{document}